\documentclass{aa}  
\usepackage{graphicx}
\usepackage[varg]{txfonts}
\usepackage{color}
\usepackage{natbib}
\bibpunct{(}{)}{;}{a}{}{,}
\begin{document}
\title{Are compact groups hostile towards faint galaxies?}
\author{Ariel Zandivarez\inst{1,2}\fnmsep\thanks{arielz@mail.oac.uncor.edu}  
\and Eugenia D\'iaz-Gim\'enez\inst{1,2} 
\and Claudia Mendes de Oliveira\inst{3}
\and Henrique Gubolin\inst{3}}

\institute{
Instituto de Astronom\'{\i}a Te\'orica y Experimental, IATE, CONICET, C\'ordoba, Argentina
\and
Observatorio Astron\'omico, Universidad Nacional de C\'ordoba, Laprida 854, X5000BGR, 
C\'ordoba, Argentina
\and 
Instituto de Astronom\'ia, Geof\'isica e Ciencias Atmosfericas, IAG, USP, Rua do Mat\~ao 1226,
S\~ao Paulo, Brazil
}
\date{Received XXX; accepted XXX}
\abstract{}
{The goal of this work is to understand whether the extreme environment of compact groups can 
affect the distribution and abundance of faint galaxies around them.
} 
{We performed an analysis of the faint galaxy population in the vicinity of compact groups
and normal groups. We built a light-cone mock galaxy catalogue constructed
from the Millennium Run Simulation II plus a semi-analytical model of galaxy formation.
We identified a sample of compact groups in the mock catalogue as well as a control sample
of normal galaxy groups and computed the projected
number density profiles of faint galaxies around the first- and the second-ranked galaxies. 
We also compared the profiles obtained from the semi-analytical galaxies in compact groups with those
obtained from observational data. In addition, we investigated whether the ranking or the luminosity of 
a galaxy is the most important parameter in the determination of the centre around which 
the clustering of faint galaxies occurs.
}
{There is no particular influence of the extreme compact group environment 
on the number of faint galaxies in such groups compared to control groups.
When selecting normal groups with separations between the 1st and 2nd ranked
galaxies similar to what is observed in compact groups, the faint galaxy projected number
density profiles in compact groups and normal groups are similar in shape and height.
We observed a similar behaviour of the population of faint galaxies in observations and simulations 
in the regions closer to the 1st and 2nd ranked galaxies.  
Finally, we find that the projected density of faint galaxies is higher around luminous galaxies, 
regardless of the ranking in the compact group.
}
{The semi-analytical approach shows that compact groups
and their surroundings do not represent a hostile enough environment to make faint galaxies to 
behave differently than in normal groups.
}

\titlerunning{Are compact groups hostile towards faint galaxies?}
\maketitle
\section{Introduction} 

Overdense regions in the universe are the most natural laboratories
to witness galaxy formation and evolution. Among them, the most extreme
environment in the universe can be found analysing compact groups (CGs) of
galaxies. These small systems of few galaxies in close proximity represent 
a key element in extragalactic astronomy for the study of galaxy interactions 
and to investigate how these lead to morphological transformation of the member galaxies
through the lifetime of the group.  

Since the construction of the first catalogues of CGs by visual
inspection \citep{rose77,hickson82}, several studies have been
performed analysing these groups and their galaxy members. 
Some of these studies have addressed topics such as internal
structures, morphologies, luminosities and environments
(e.g., \citealt{hickson84,mamon86,hickrood88,mendes91,prandoni94,
kelm04}), in an attempt to unveil their formation scenario. 
During years, evidence has been gathered to clarify whether they
are recently formed systems that are about to coalesce into a single
galaxy \citep{hickrood88}, or transient unbound cores of looser groups
\citep{tov01}, or just simple chance alignments of galaxies along the 
line of sight within larger systems \citep{walke89,hernquist95}.

Since most of the famous CGs were identified relying
on a visual inspection (following Hickson selection criterion - 
\citealt{hickson82}), the intrinsic nature of CGs has always been 
challenging, prompting scientist to perform different kinds of 
numerical tests to quantify their reliability. 
Among those, the studies performed using numerical 
simulations plus semi-analytic models have proven to be a very 
important tool to improve our understanding of these groups \citep{mcconnachie08,diaz10}.
For instance, such studies confirm that roughly 70 per cent of 
CGs identified using Hickson's criterion can be considered physically 
dense groups and that the Hickson's sample is incomplete \citep{diaz10}.

Even though many efforts have been devoted to study CGs, their
members and the signs of interaction as a tool to understand all
the relevant processes involved in galaxy evolution, most of them have
overlooked the impact of these extremely dense regions on numbers and distributions of 
the faint galaxies.
These galaxies are a very important population in galaxy systems 
(e.g., \citealt{gonzalez06}),   
and given the high density and the low velocity dispersion of CGs,
they are very suitable to analyse the effects of interaction-inductive environments on 
the faint galaxy structure and evolution. 
It is usually expected that such dense environment of CGs
can drastically affect the evolution of their brightest members, but
are these environments hostile enough to alter the faint galaxy
population distribution as well? 
Performing such studies using observations has been relegated mainly due 
to two different aspects: first, by definition of CGs, galaxies are 
considered as members of these systems only if their magnitudes are 
within a 3-magnitude range from the brightest member, 
therefore fainter galaxies are just ignored in the samples; 
second, it could be difficult to detect these faint objects 
and actually only few studies have been carried out analysing 
this population in individual CGs (e.g., 
\citealt{ribeiro94,zablu98,amram04,campos04,carrasco06,krusch06,darocha11,konstantopoulos13}).
Hence, more evidence is needed to understand the relevance of this 
particular environment on faint galaxies in/around CGs.

Given that these types of studies require the use of catalogues which contain a 
large population of faint galaxies as well as spectroscopic 
information to ensure proximity to the environment of CGs, 
an optimal way to approach the problem is performing this analysis using synthetic 
catalogues constructed from numerical simulations plus semi-analytic models 
of galaxy formation. 
Recently, a high resolution N-body numerical simulation has been released, the
Millennium Run Simulation II \citep{mII},  
perfect for resolving dwarf galaxies using semi-analytic recipes.
A particular set of recipes was already applied to this simulation, 
\citealt{guo11}, producing a very suitable sample of synthetic galaxies 
that has been made available for the astronomical community.
The semi-analytic model has been tuned to reproduce the $z=0$ stellar
mass function and luminosity function, making it a suitable tool to 
understand the evolution of faint galaxies.
Therefore, in this work we will use this available tool to study,
from the semi-analytical point of view, whether the population of 
faint galaxies in/around CGs is affected by the environment 
of these systems.

The layout of this paper is as follows: in section 2, we describe
the N-body simulation and the semi-analytic model of galaxy formation
used to build the mock catalogue. In section 3 we describe the 
different procedures adopted for the CG identification as
well as the construction of a control sample of normal groups. 
In section 4 we construct the sample of faint galaxies around the
galaxy systems and compute the projected number density profiles taking
into account the properties of faint galaxies and those of the CGs they inhabit. 
In section 5 we perform a 
comparison between the projected number density profiles obtained from mock 
catalogues with those obtained for a sample of CGs 
identified from observations. Finally, in section 6 we summarise 
our results.

\section{The mock galaxy catalogue}
We built a light-cone mock catalogue using a simulated set of galaxies extracted
from the \cite{guo11} semi-analytic model of galaxy formation applied on 
top of the Millennium Run Simulation II \citep{mII}. 

\subsection{The N-body simulation}
The Millennium Run Simulation II is a cosmological Tree-Particle-Mesh 
\citep{xu95} N-body Simulation, which evolves 
10 billion ($2160^3$) dark matter particles in a 100 $h^{-1} \ Mpc$ periodic 
box, using a comoving softening length of 1 $h^{-1} \ kpc$ \citep{mII}. The cosmological 
parameters of this simulation are consistent with WMAP1 data \citep{spergel03}, 
i.e., a flat cosmological model with a 
non-vanishing cosmological constant ($\Lambda$$CDM$): $\Omega_m$ =0.25, 
$\Omega_b$=0.045, $\Omega_{\Lambda}$=0.75, $\sigma_8$=0.9, $n$=1 and $h$=0.73. 
The simulation was started at $z$=127, with the particles initially positioned 
in a glass-like distribution according to the $\Lambda$$CDM$ primordial 
density fluctuation power spectrum. The mass resolution is 125 times better
than the obtained in the Millennium Run Simulation I \citep{springel05}, i.e, 
each particles of mass has $6.9\times10^6 h^{-1} \ M_{\odot}$. 
With this resolution, halos of typical dwarf
spheroids are resolved and halos similar to the mass of our Milky Way have
hundreds of thousands of particles \citep{mII}.

It is well known that WMAP7 \citep{komatsu11} yielded different cosmological parameters
than the ones used here (from WMAP1). Therefore,
one may argue that the studies carried out in the present simulation could produce results 
that do not agree with the current cosmological model. However, \cite{guo13} have 
demonstrated that the abundance and clustering of dark halos and galaxy properties, 
including clustering, in WMAP7 are very similar to those found in WMAP1 for $z\leq3$, 
which is far inside the redshift range of interest in this work (see Sect.~\ref{mock}).

\subsection{The semi-analytic model}
To obtain a simulated galaxy set we adopted the \cite{guo11} semi-analytic 
model. This particular model fixed several open issues present in some 
of its predecessors, such as the efficiency of supernova feedback and 
the fit of the stellar mass function of galaxies at low redshifts.
\cite{guo11} also introduced a more realistic treatment 
of satellite galaxy evolution and of mergers, allowing satellites to continue 
forming stars for a longer period of time and reducing the satellite 
excessively rapid reddening. The model also includes a treatment of the 
tidal disruption of satellite galaxies. 
Compared to previous versions
of the semi-analytical models, \citeauthor{guo11} model has lower number of galaxies 
than its predecessors, at any redshifts and in any environment. This is the result of
a stronger stellar feedback that reduces the number of low-mass galaxies, and a model
of stellar stripping, which contributes to reduce the number of intermediate 
to low mass galaxies \citep{vulcani14}. 

This model produces a complete sample when considering galaxies with 
stellar masses larger than $\sim 10^{6.4} h^{-1} \ M_{\odot}$. This implies
that the galaxy sample is almost complete down to an absolute magnitude in
the $r_{SDSS}$-band of -11.

\subsection{Mock catalogue construction}
\label{mock}
To construct a mock galaxy catalogue we followed a similar procedure to
 that described in \cite{fof14}. Basically, the procedure is as follows:
\begin{itemize}
\item We locate a virtual observer at zero redshift and find the galaxies 
that lie on the observer's backward light-cone. The catalogue is constructed 
by adding shells taken from different snapshots corresponding to the epoch of 
the lookback time at their corresponding distance \citep{diaz02,henriques12,wang12}.
\item Given the requirements of our work with CGs, we introduce a maximum 
redshift of $0.5$.
\item Due to the limited size of the simulation box, $100 \ h^{-1} \ Mpc$, 
we use the periodicity of the box and build a super-box to reach the desired 
maximum distance.
\item The cosmological redshift ($z_c$) is obtained from the comoving distance 
of the galaxies in the super-box and the distorted or spectroscopic redshift ($z_s$) 
is computed considering the peculiar velocities of the galaxies in the 
radial direction ($v_p$), i.e., $z_s=(1+z_c)(1+v_p/c)-1$,
 where $c$ is the velocity of the light.
\item To avoid discreteness of the galaxy magnitudes due to the size of the shells, 
we interpolate the absolute magnitudes among two consecutive shells, 
according to their distance to the shell edges.
\item We use the prescription of \cite{fof14} to avoid the problems of repeated or 
missing galaxies in the boundaries of two consecutive shells due to the proper 
movements of the galaxies from one snapshot to the next.
\item We compute k-corrections for each galaxy using an iterative procedure based on 
the prescriptions given by \cite{chilinga10}.
\end{itemize}
The final mock catalogue comprises $775439$ galaxies down to an apparent magnitude limit $r_{lim}=16.3$ within a solid angle of $4\pi$. 
We also selected a sample of galaxies within the same volume,
 but having apparent magnitude limit of 21, 
which will be used to select the faint neighbours. 
This new sample has $\sim 140$~million galaxies. 
\begin{figure}
\begin{center}
\includegraphics[width=8cm]{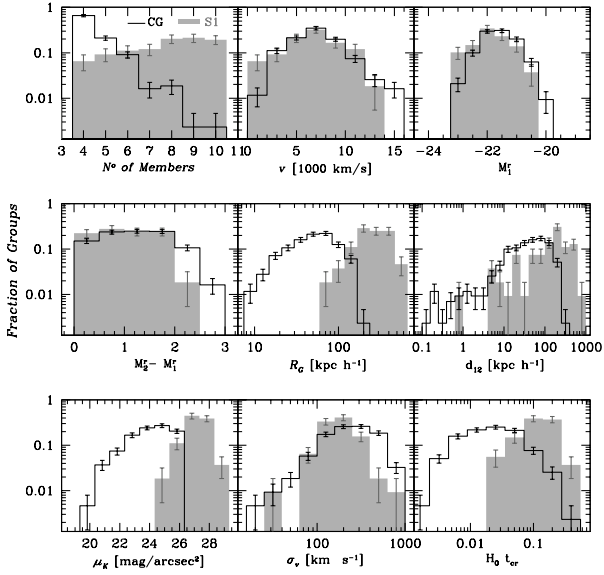}
\caption{
Distributions of observable properties of Compact and Control Groups. 
Number of members in a 3-magnitude range (\emph{top-left
panel}), radial velocity of the groups (\emph{top-centre panel}), 
rest frame r-band absolute magnitude of the brightest galaxy member (\emph{top-right panel}),
difference in absolute magnitude between the brightest and the 
second brightest galaxy of the groups (\emph{middle-left panel}),
projected radius of the minimum circle that encloses the galaxy members (\emph{middle-centre panel}), 
projected distance between the brightest and second brightest galaxies (\emph{middle-right panel}), 
mean group surface brightness (\emph{bottom-left panel}),
radial velocity dispersion of the groups (\emph{bottom-centre panel}), and
dimensionless crossing time (\emph{bottom-right panel}).
\emph{Black empty histograms} correspond to Compact Groups (CGs), 
while \emph{grey histograms} correspond to Control Groups (S1). 
Error bars correspond to Poisson errors.}
\label{f1}
\end{center}
\end{figure}

\section{Selecting the samples of groups}

\subsection{Compact group sample}
We defined a mock velocity-filtered compact group sample
following the prescriptions of \cite{diaz10} and \cite{diaz12}. 
Briefly, the automated searching algorithm
defines as compact groups those that satisfy the following criteria:
\begin{itemize}
\item $4 \le N \le 10$ (population)
\item $\mu_r \le 26.33 \, \rm mag \, arcsec^{-2}$ (compactness)
\item $\Theta_N > 3\Theta_G$ (isolation)
\item $r_{brightest} \le  r_{lim}-3=13.3$ (flux limit)
\item $|v_i -\langle v \rangle| \le 1000 \ \rm km/s $ (velocity filtering)
\end{itemize}
where N is the total number of galaxies whose r-band magnitude
satisfies $r < r_{brightest} + 3$, and $r_{brightest}$ is the apparent magnitude
of the brightest galaxy of the group; $\mu_r$ is the mean r-band surface
brightness, averaged over the smallest circle circumscribing
the galaxy centres; $\Theta_G$ is the angular diameter of the smallest circumscribed
circle; $\Theta_N$ is the angular diameter of the largest
concentric circle that contains no other galaxies within the considered
magnitude range or brighter; $v_i$ is the radial velocity of each galaxy member, and $\langle v \rangle$ is the median of the radial velocity of the members. 
The compactness and flux limit criteria are set to match the identification 
in the 2MASS catalogue performed in \cite{diaz12}.

As in previous works, we have considered that mock
galaxies that are close in projection on the plane of the mock sky
would be blended by observers if their angular separation is less
than the sum of their angular half-light radii. 
Therefore, only groups with more than 4 members after blending galaxies 
survive.

This algorithm identifies $431$ CGs.

\begin{figure}
\begin{center}
\includegraphics[width=8cm]{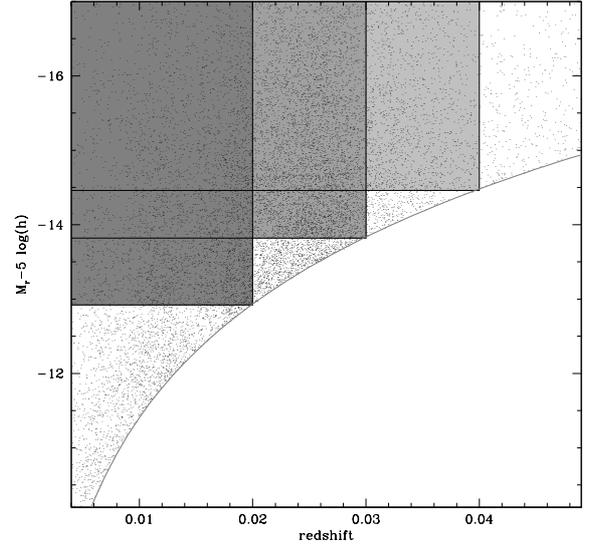}
\caption{Absolute magnitude of faint galaxies selected within the composite cylinder 
vs. redshift. The lower envelope represents the apparent magnitude limit of the
catalogue ($r=21$).
\emph{Grey boxes} represent three volume limited samples defined as follows: 
$-17\le\rm M_r-5 \, log(h)\le-12.9$ \& $z\le0.02$, 
$-17\le\rm M_r-5 \, log(h)\le-13.8$ \& $z\le0.03$, 
$-17\le\rm M_r-5 \, log(h)\le-14.5$ \& $z\le0.04$.
}
\label{f4}
\end{center}
\end{figure}
\begin{figure}
\begin{center}
\includegraphics[width=6cm]{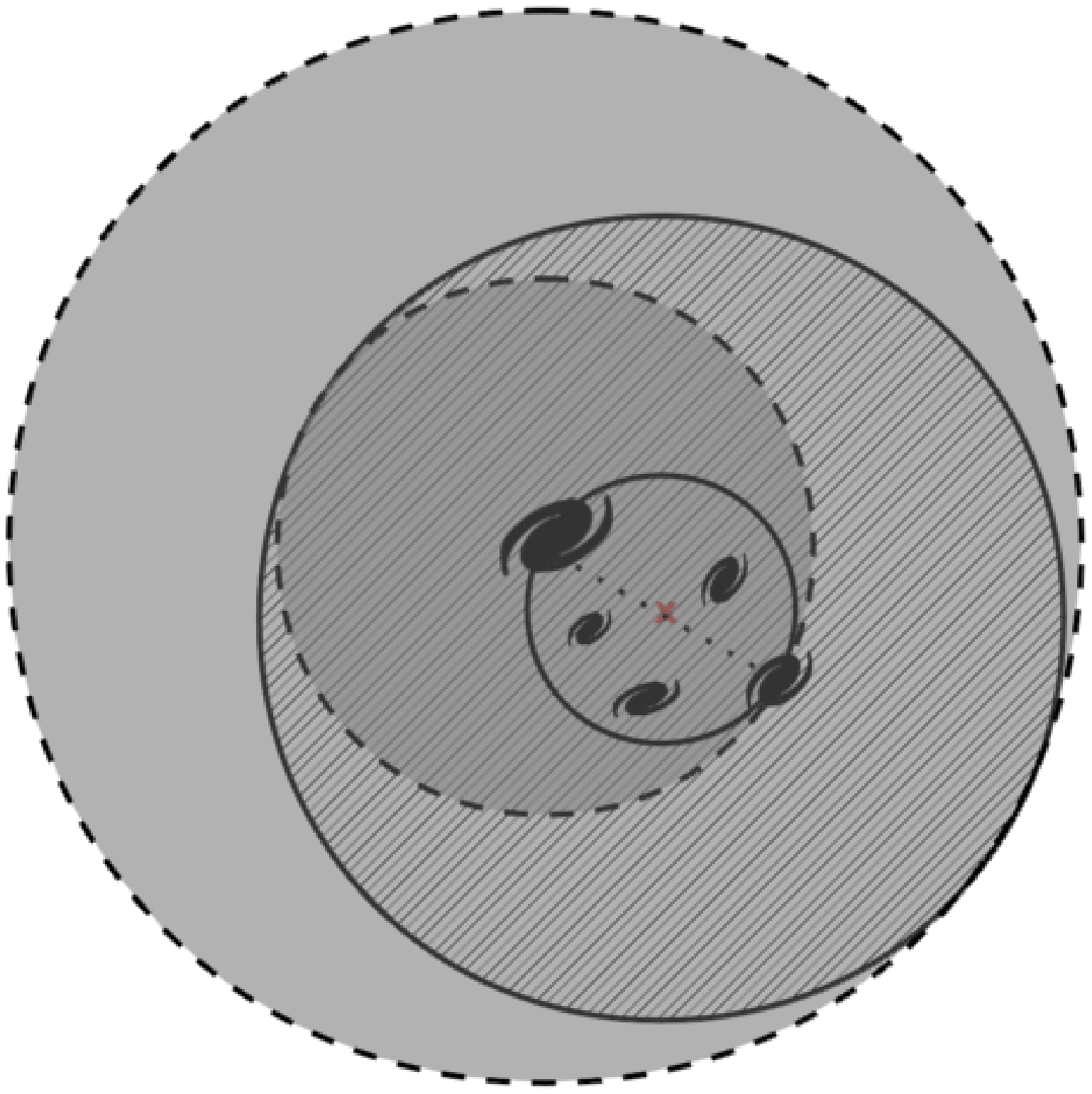}
\includegraphics[width=4.5cm]{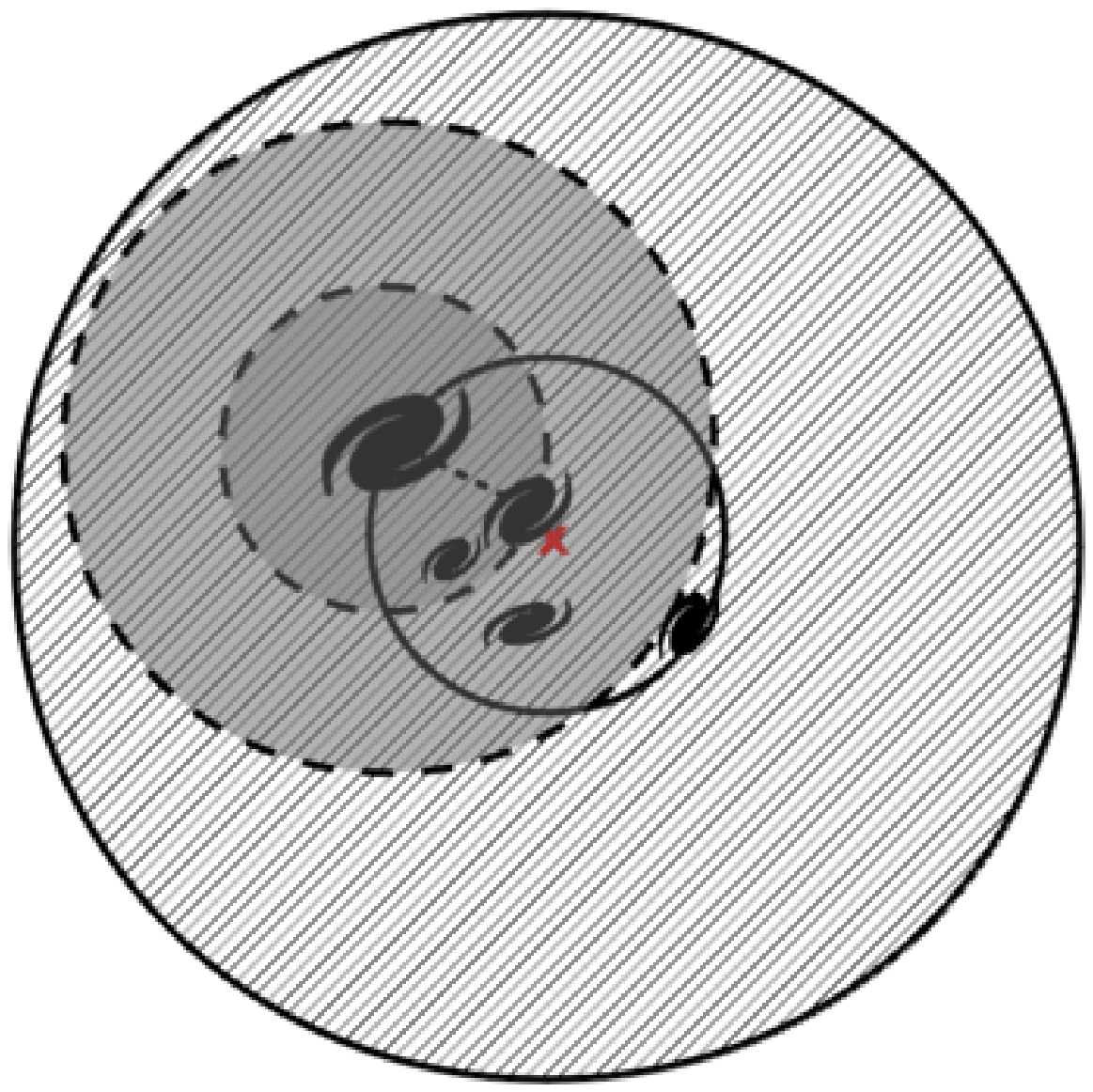}
\caption{Illustrations of two CGs and the projected cylinders around 
the centre of the minimum circle (where faint galaxies were selected), 
and around the 1st(or 2nd) ranked galaxies (where the density profiles are measured).
\emph{Solid lines} show the minimum circle and 3 times the minimum circle. 
\emph{Dashed lines} show the circles centred in the first ranked galaxies and having 
radii of $d_{12}$ and  $2 \, d_{12}$. 
Galaxies shown in this figure represent only those considered as compact group members 
(within a 3-mag range from the brightest).
\emph{Upper plot} shows an extreme case where the centre of the profile is on the edge of 
the minimum circle that encloses the CG galaxy members and the separation between the 
two brightest galaxies is the diameter of the minimum circle. In this extreme case, the dashed line with radius
$d_{12}$ is the largest circle around the 1st or 2nd ranked galaxy that falls entirely 
within the cylinder defined for the search of faint galaxies (solid line).
\emph{Lower plot} shows a generic case where the centre of the profile is also on the edge of 
the minimum circle that encloses the CG galaxy members, but now the separation between the 
two brightest galaxies is smaller than the radius of the minimum circle. 
In this case,  it is possible to go farther than $2 \, d_{12}$ and still be complete in
the sample of faint galaxies.
}
\label{f2}
\end{center}
\end{figure}
\subsection{Control group sample}
To build a group sample to compare the results obtained for CGs, 
we identified galaxy groups in the same mock galaxy catalogue. The
identification was performed using a Friends-of-Friends algorithm similar to that
developed by \cite{huchra82} to identify galaxy systems in redshift space for a 
flux-limited catalogue. The algorithm links galaxies that share common neighbours,
i.e, pairs of galaxies with projected separations smaller than $D_0$ and radial
velocities differences lesser than $V_0$.
Following the prescriptions of \cite{fof14}, we used a radial linking length of 
$V_0=130\ km/s$ and a transversal linking length, $D_0$, defined by a contour overdensity 
contrast of $\delta\rho/\rho=433$ (see Eq. 4 of \citealt{huchra82}). 
This value of $\delta\rho/\rho$ is adopted
since it is expected that galaxies are more concentrated than dark matter 
\citep{eke04,berlind06}, therefore we should use a higher density contrast than 
that usually adopted in dark matter simulations, between 150-200 
(see Appendix B of \citealt{fof14} for details). 
 
Since we are dealing with a flux limited sample of galaxies, both linking lengths
have to be weighted by a factor, $R$, to take into account the variation 
of the sampling of the 
luminosity function produced by the different distances of the groups to the observers.
That factor $R$ was proposed by \cite{huchra82} and is computed using the
galaxy luminosity function following the equation
\begin{equation}
\nonumber
R=\left[ \frac{\int_{-\infty}^{M_{12}} \phi(M) dM}{\int_{-\infty}^{M_{lim}} \phi(M) dM} \right]^{-1/3}
\end{equation}
where $M_{lim}=-13.7$, and $M_{12}=r_{lim}-25-5\log(dL_{12})$ with $dL_{12}$ the mean
luminosity distance for the galaxy pair.
Therefore, we used the information from the semi-analytic model
to compute the luminosity function for the $z=0$ snapshot of the simulation.
Using a Levenberg-Marquardt method, we fitted a double-Schechter function 
to the distribution of rest frame $r_{SDSS}$ absolute magnitudes:
\begin{equation}
\nonumber
\phi(L)=\frac{1}{L^{\ast}}\exp\left( -\frac{L}{L^{\ast}} \right) \left[ \phi_1 \left( \frac{L}{L^{\ast}} \right)^{\alpha_1} + \phi_2 \left( \frac{L}{L^{\ast}} \right)^{\alpha_2} \right]
\end{equation}
obtaining as best fitted parameters: $M^{\ast}-5\log(h)=-20.3\pm0.1$, $\phi_1=0.0156\pm0.0003$, 
$\alpha_1=-0.06\pm0.02$, $\phi_2=0.0062\pm0.0003$ and $\alpha_2=-1.41\pm0.02$.

Using the all-sky mock catalogue, this algorithm produced a sample of $1897$ 
groups with 10 or more galaxy members with magnitudes brighter than $r_{SDSS}$-band 
of 16.3 and up to redshift 0.1

We imposed the following constraints on the identified group sample 
to select the sample of control groups:
\begin{figure*}
\begin{center}
\includegraphics[width=11cm]{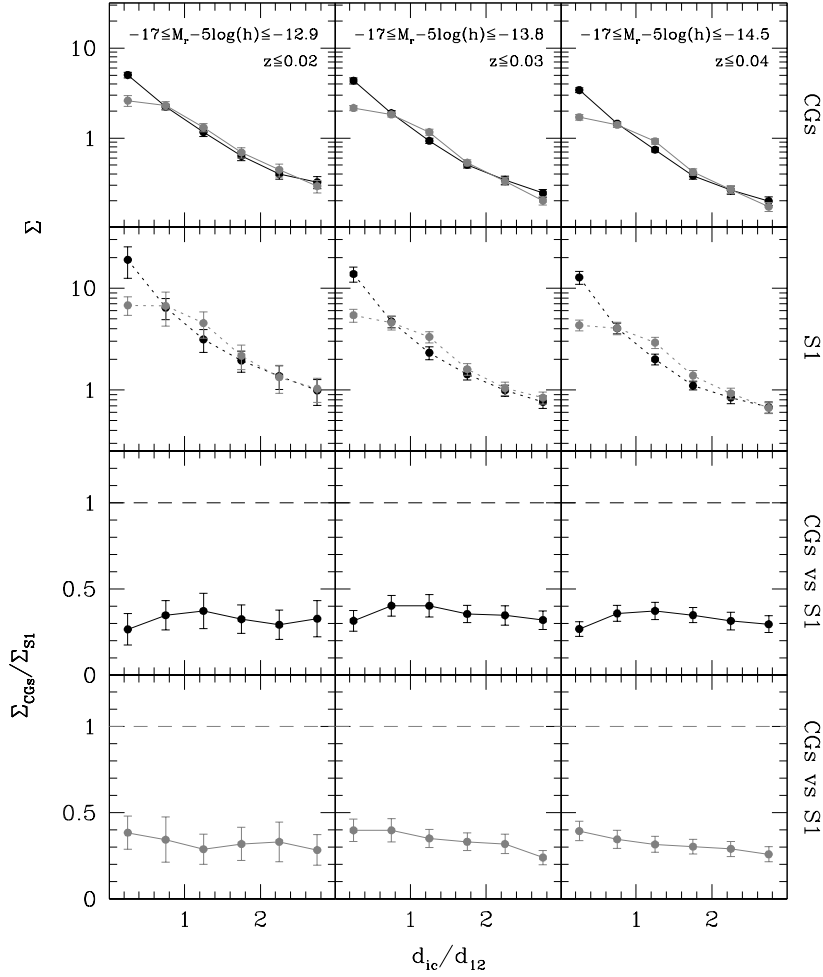}
\caption{Projected number density profiles of faint galaxies around 
the first ranked (\emph{black}) and second ranked galaxies (\emph{grey}), split into 
three volume-limited samples (columns from left to right).  
\emph{Solid lines} correspond to the profiles in compact groups (CGS),
 while \emph{dotted lines} are around the control group sample (S1).
Error bars are the standard deviations computed with 100 bootstraps.
\emph{Bottom panels} show the ratios of the profiles around CGs and S1s.
Errors are computed by error propagation.
}
\label{f5}
\end{center}
\end{figure*}

\begin{itemize}
\item $r_{brightest}<13.3$. This limit was imposed to match the flux limit criterion in the 
identification of CGs. This restriction ensures that all the groups have the same 
chance of including all the existing galaxies within a 3-magnitude range from the brightest. 
The sample that remains comprises 741 galaxy groups.
\item $4\le N_3^g \le10$, where ($N_3^g$) is the number of members in a 3-magnitude range 
from the brightest. This constraint was included to mimic the galaxy population 
allowed in the CGs sample. This limit restrict the group sample to 315 systems. 
\item We also discarded those galaxy groups that 
share galaxies with the Compact Groups of our sample.
The remaining sample comprises 231 galaxy groups.
\item Finally, we excluded from our control group sample those 
with a bright galaxy (brighter than
the brightest galaxy of the FoF group) inside the cylinder used to compute the 
projected density profiles 
(see next section for further details about the cylinder definition). 
This restriction produces a final control group sample of 108 galaxy groups. 
\end{itemize}
The sample of control groups is referred as S1 sample in the figures that follows. 
It is worth mentioning that the different algorithms used to identify the compact groups and 
control groups mean that the two samples do not completely overlap. I.e. not all compact groups 
meet the requirements for inclusion in the sample of control groups (prior to the exclusion of 
groups that are in the compact group sample, obviously), and vice versa.

In Fig.~\ref{f1} we show a comparison between the observable group properties of the
CG sample (\emph{black empty histograms}) and the S1 sample 
(\emph{grey histograms}). All the properties have been computed using only the members 
within a 3 magnitude range from the brightest galaxy. 
We show properties that are commonly derived when studying compact groups, although 
some of them are barely defined for normal groups 
(for instance: the radius of the minimum circle that encloses all the galaxy members).
From this comparison, it can be seen that both samples span similar ranges of distances,
 magnitudes of the brightest galaxy and radial velocity dispersions. 
As expected, normal groups show larger membership, projected size, 
projected separation between their two brightest galaxies, and fainter surface brightness. 
They also show larger crossing times, computed as 
\[
H_0 \, t_{\rm cr} = H_0 \, \frac{\langle d_{ij}^{3D}\rangle}{\sigma_{3D}} =
\frac{100 \, \pi}{2 \sqrt{3}} \,h\, \frac{ \langle
  d_{ij}\rangle}{\sigma_v }  \ , \] 
where $\langle d_{ij}\rangle$ is the median of
the inter-galaxy projected separations in $h^{-1} \,\rm Mpc$.

\begin{figure}
\begin{center}
\includegraphics[width=8cm]{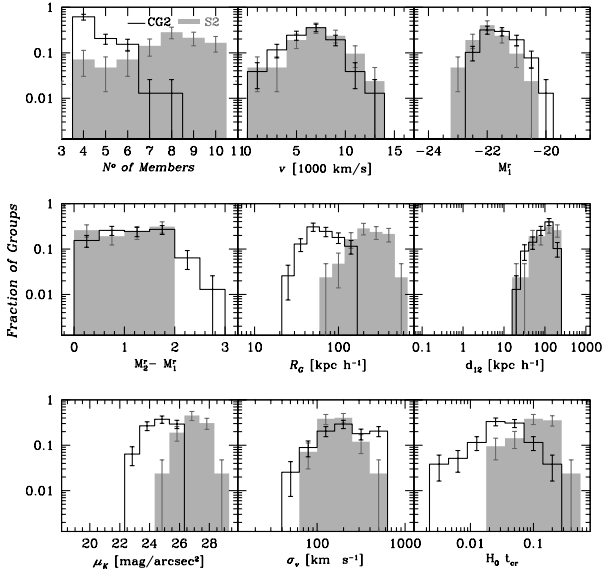}
\caption{Same as Fig.~\ref{f1} for mock CGs2 
(\emph{black empty histograms}) and control groups S2 (\emph{grey histograms}) 
after matching their angular distances between the 1st and 2nd ranked galaxies.
Error bars correspond to Poisson errors.
  }
\label{f6}
\end{center}
\end{figure}

\section{Projected density profiles of faint galaxies}
\subsection{Faint galaxies in/around groups}
\label{faints}
We consider as "faint galaxies" those galaxies that were not included as group members 
given the population criterion considered by Hickson, but that inhabit the same region in the sky.
Faint galaxies around the groups were selected from the catalogue of galaxies 
brighter than $r=21$ defined at the end of Sect.~\ref{mock} as follows. 

For CGs, we define a cylinder for searching faint galaxies centred in projection at the
centre of the minimum circle that encloses the CG members. 
Faint galaxies were selected with the following criteria:
\begin{itemize}
\item $\rm M_{r_i} -5\, \log{(h)} \ge -17$
\item $\phi_i < 3 \Theta_G/2$
\item $|v_i -\langle v \rangle| \le 1000 \, \rm km/s $ 
\end{itemize}
where $M_{r_i}$ is the $r_{SDSS}$-band rest frame absolute magnitude, 
$\phi_i$ is the angular distance of the galaxy $i$ to the centre of the minimum circle 
that encloses the CG members, and $v_i$ is its radial velocity.
The upper limit $-17$ is adopted to avoid group members linked by the identification algorithm
 being included in the sample of faint neighbours. 
It can be seen in the top right panel of Fig.~\ref{f1} that the faintest first-ranked
galaxy is brighter than $-20$, therefore the group members (which by definition span a 3-magnitude 
range from the brightest) are always brighter than $-17$. 

Similar criteria were used to select faint galaxies around control groups. 
In this case, the cylinder is centred in projection at the position of
the brightest galaxy of the system. Then, faint galaxies satisfy:
\begin{itemize}
\item $\rm M_{r_i} -5\, \log{(h)} \ge -17$
\item $\psi_i < 4 \Theta_{12}$
\item $|v_i -\langle v \rangle| \le \Delta v $ 
\end{itemize}
$\psi_i$ is the angular distance of the galaxy $i$ to the brightest galaxy 
of the group, $\Theta_{12}$ is the angular distance between the brightest and 
the second brightest galaxy of the group, 
and $\Delta v$ is the maximum difference in radial velocity 
between the FoF group members and the median of the group. 

Notice that the definition of the cylinders in which the faint galaxies 
are considered differs from one sample to the other.
For CGs, the cylinder is straightforwardly defined based on the criteria of 
isolation of CGs. By definition, there are no other galaxies brighter 
than $r_{brightest}+3$ within 
three times the angular radius and $ 2000 \, \rm km/s $ in the line of sight, 
although it may contain fainter galaxies that 
are ignored in the properties of the CGs and do not affect neither 
the isolation nor the compactness criteria.
For control groups, since they are identified with a FoF algorithm, 
the shape of the groups varies from group to group. 
In the line of sight direction, we used the members originally linked 
by the FoF algorithm ($r<16.3$) to determine the length of the cylinder, 
which is always shorter than $2000 \, \rm km/s $.
We chose to define the projected cylinder centred on the first ranked galaxy 
of the group, and used the separation between the first and second ranked 
galaxies as a proxy for the projected radius of the cylinder.
 
\begin{figure*}
\begin{center}
\includegraphics[width=11cm]{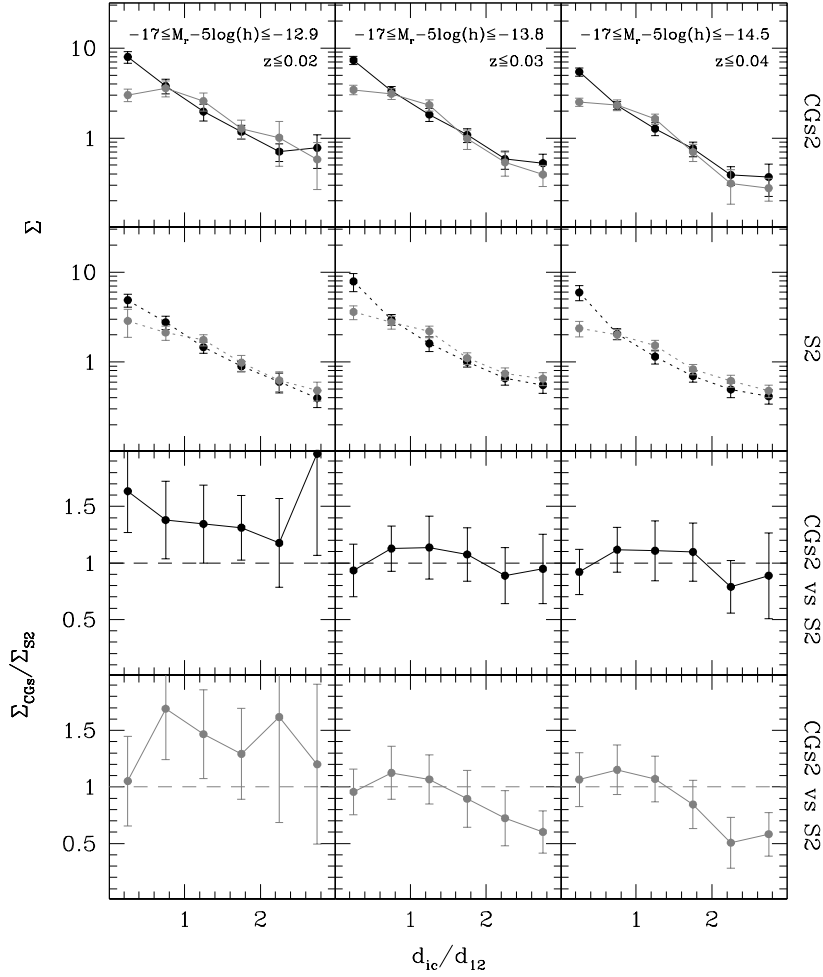}
\caption{Projected number density profiles of faint galaxies around 
the first ranked (\emph{black}) and second ranked galaxies (\emph{grey}), split into 
three volume-limited samples (columns from left to right).  
\emph{Solid lines} correspond to the profiles in compact groups (CGs2),
 while \emph{dotted lines} are around the control group sample (S2). 
Error bars are the standard deviations computed with 100 bootstraps.
\emph{Bottom panels} show the ratios of the profiles around CGs2 and S2s.
Errors are computed by error propagation.
}
\label{f7}
\end{center}
\end{figure*}

\subsection{Number density profiles: volume limited samples}
\label{profiles}
Using a flux limited catalogue implies that 
groups at different redshifts are inhabited by different 
populations of galaxies in terms of their intrinsic luminosities. 

Fig.~\ref{f4} shows the absolute magnitudes of galaxies selected 
within the group cylinders vs. their redshifts. The lower envelope reproduces the 
apparent magnitude limit $r=21$. To avoid possible biases related with incomplete 
sampling of galaxies in terms of luminosity, we defined three volume limited samples.
The selected samples of faint galaxies lie on the grey boxes of Fig.~\ref{f4}, and 
are defined by the following criteria:
\begin{itemize}
\item $\rm -17 \le M_{r_i} - 5 \, \rm log(h) \le -12.9$ \& $z_{cm}\le 0.02$, 
\item $\rm -17 \le M_{r_i} - 5 \, \rm log(h) \le -13.8$ \& $z_{cm}\le 0.03$, 
\item $\rm -17 \le M_{r_i} - 5 \, \rm log(h) \le -14.5$ \& $z_{cm}\le 0.04$. 
\end{itemize}
where $\rm M_{r_i}-5 \, \rm log(h)$ is the galaxy absolute magnitude in the $r_{SDSS}$ band and $z_{cm}$ is
the median redshift of the system.

Therefore, we computed the projected number density profiles of faint galaxies that belong 
to each of those complete subsamples around the 
first and second ranked galaxies of the groups, for CGs and control groups,. 
To obtain statistically significant results, we built a composite group by 
stacking the groups normalised to a characteristic size. We adopted the projected
distances between the first and second ranked galaxies, $d_{12}$, as a normalisation size. 
After computing the number of faint galaxies within the areas of the projected rings
around the first and second ranked galaxies, 
we divided the projected density by the number of groups that contributed to each bin 
of normalised distance.

It is worth mentioning here the influence of our choice of cylinders to 
look for the faint galaxies. 
In normal groups, we selected faint galaxies up to 4 times 
the normalisation factor, $d_{12}$, around the first ranked galaxy. 
Therefore, we can assure that all groups contribute up to 4 times the $d_{12}$ 
around the first ranked galaxy, and also we can assure that all groups 
contribute at least up to 3 times the $d_{12}$ around the second ranked galaxy. 
In both cases, we will only be interested in the profiles up to 3 times 
the normalisation distance. 

In compact groups, however, the cylinder in which the faint galaxies are selected 
was defined based on the CG criteria: 
centred in the centre of the minimum circle and spanning up to 3 times 
the size of that circle. 
But, given that the profiles will be centred in the 1st- and 2nd-ranked galaxies, some 
considerations have to be taken into account. 
In the worst scenario shown in the \emph{upper plot} of Fig.~\ref{f2}, 
the $d_{12}$ may be as large as the diameter of the minimum circle 
($2 \, R_G$, where $R_G$ is the projected radius of the minimum circle).
In such cases, the projected rings around the first ranked galaxy 
will be complete only up to a radius of $d_{12}$, therefore for those groups 
we do not take into account the contribution of galaxies whose normalised
distances to the first ranked galaxies are larger than $1$. 
The same is valid when the 2nd-ranked galaxy is the centre of the profile.
As the centre of the density profile (1st or 2nd ranked galaxy) is closer to 
the centre of the minimum circle or the separation between the 1st and 2nd galaxies
is shorter, the contribution of faint galaxies will span out 
to larger normalised distances, 
as it is shown in the \emph{lower plot} of Fig.~\ref{f2}. 
Also, the number of groups that contribute in each bin of normalised distance 
varies accordingly. 

Given that we are computing the projected number density profiles up to 3 times the 
normalisation size, and that we are carefully taking into account the proper normalisation
of the group contributions, the different definitions of cylinders for compact and 
control groups do not affect the resulting projected number density profiles. 

Figure~\ref{f5} shows the projected number density profiles around the first 
(\emph{black}) and second (\emph{grey}) ranked galaxies, 
for CGs (\emph{solid lines}) and control groups (\emph{dashed lines}). Error bars are 
computed using 100 bootstrap resamplings. Each column corresponds to each of the 
different volume limited samples defined above. 
In the bottom panels of this figure, we show  
the ratio between the profiles of CGs and S1 around the first and second ranked galaxy.
Error bars are computed by error propagation. 
We find that the number density of faint galaxies is larger around normal groups 
than around CGs. However, since the ratios are almost constant, 
the way those galaxies are distributed are very similar.
In both, CGs and S1 groups, faint galaxies mainly cluster 
around the brightest galaxy of the group which is reflected in a more cuspy profile around 
the 1st ranked galaxy. As expected, the second brightest galaxy also gathers
faint galaxies, but the effect is not as important as around the first ranked galaxy. 
In fact, the density of faint galaxies is determined by the brightness of the galaxies 
taken as centres rather than the ranking they have inside the group (see Appendix~\ref{appen1}).

At this point we analysed whether the differences in the sizes of the 
normalisation parameters in CGs and S1 could lead to the difference 
we found in the projected number density of faint galaxies. 

Therefore, we defined a subsample of CGs and a subsample of control groups, 
in order to avoid any dependence of the 
profiles on the separation of the two brightest galaxies:
we restricted both samples to those groups whose angular distance between
the two brightest galaxies is $1.5 \ {\rm arcmin} \le \Theta_{12} \le 10 \ {\rm arcmin}$ 
and matched the normalised $\Theta_{12}$ distributions. 
We refer to these subsamples as CGs2 and S2. 
The CGs2 sample comprises 78 compact groups, while the S2 sample 
comprises 42 normal groups. 
Figure~\ref{f6} shows the distribution of properties of CGs2 and S2 groups. 
From the comparison with Fig.~\ref{f1}, it can be seen that 
compact groups with the brightest surface brightness have been 
excluded after restricting the separations between their two brightest galaxies, 
anyway, they have still brighter surface brightness than the control groups 
and have smaller projected radii. 

We repeated the calculation of the projected number density profiles by splitting the 
faint galaxies into the three volume-limited samples defined above. 
The volume-limited subsamples comprise 30(12), 70(34) and 76(41) CGs2(S2), from 
the closest to the deeper volume respectively. 
Figure~\ref{f7} shows the profiles
around these subsamples of groups that span the same range of normalising distances. 

The projected number densities of faint galaxies in these new samples are different
from those observed in Fig.~\ref{f5}. It can be seen that the profiles of faint 
galaxies are now very similar in CGs and control groups
(CGs2 vs S2). Taking into account the corresponding errors 
(which are large in the smallest volume limited sample given the low number 
of groups involved), we observe that the
ratios among the projected number density profiles behave almost constant and equal to
unity in the three volume limited samples around the first ranked galaxy 
(black points/lines).  
The ratios around the second ranked galaxies also behave similarly to the former 
ratios, although CGs are underdense in the outskirts of the cylinders for the two deeper 
volume limited samples. 

\section{Comparison with observations}
\label{sdss}

\begin{figure}
\begin{center}
\includegraphics[width=8cm]{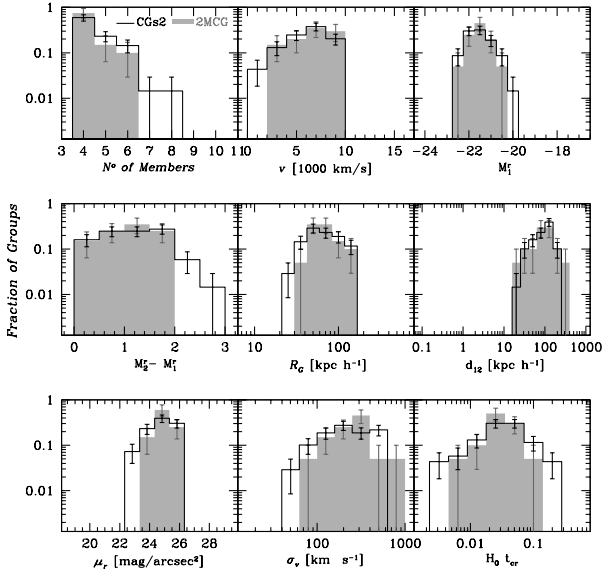}
\caption{Same as Fig.~\ref{f1} for mock CGs2 
(\emph{black empty histograms}) and 2MASS compact
groups with their faint galaxies extracted from the Sloan Digital 
Sky Survey (\emph{grey histograms}). Error bars correspond to Poisson errors.
In this figure, the mock CGs2 sample comprises the remaining groups after taking into 
account the observational incompleteness. }
\label{f8}
\end{center}
\end{figure}
\begin{figure}
\begin{center}
\includegraphics[width=7.0cm]{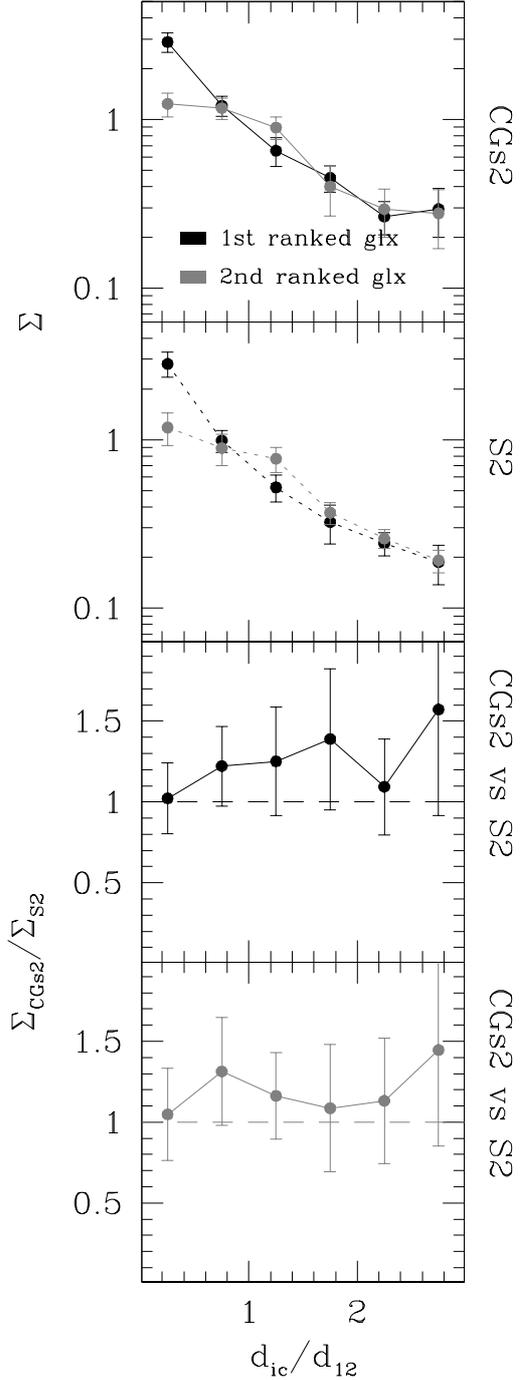}
\caption{Projected number density profiles of faint galaxies around the first ranked
(\emph{black}) and second ranked galaxy of the groups (\emph{grey}).  
The \emph{upper panel} shows the profiles 
for the mock compact group (CGs2) sample while the \emph{second} panel shows the profiles
for the control groups (S2) shown in Fig.~\ref{f6}.
Faint galaxies are selected having $r_{brightest}+3 < r_i < 17.77$
Error bars are the standard deviations computed with 100 bootstraps.
\emph{Bottom panels} show the ratios between the profiles around 
CGs2 and the profiles around S2. Errors are computed by error propagation.
}
\label{ff}
\end{center}
\end{figure}

We used the sample of 2MCG identified on the 2MASS catalogue by \cite{diaz12} 
and looked for faint neighbours in/around CGs from the spectroscopic data of the
Sloan Digital Sky Survey Data Release 9 (SDSS DR9, \citealt{dr9}) 
having apparent magnitudes brighter than $r_{lim}=17.77$.
From the original 85 2MCGs, 45 of them lie on the SDSS area.
For the purposes of this work, we have restricted the sample to CGs whose brightest galaxy 
is brighter than $r=13.27$, also, we have checked that all the selected CGs fulfil
the CG criteria in the r-band (they were originally identified as compact groups in the $K_s$-band). 
These conditions reduced the sample to 30 CGs. 

Due to the flux limit of this catalogue, faint neighbours are selected as being 
brighter than $17.77$ (and excluding the members). 
Using the Catalog Archive Server Jobs System (CasJobs) of SDSS\footnote{http://skyserver.sdss3.org/casjobs/}, 
we found that 20 of the 30 CGs have $133$ faint neighbours within 
$3 \Theta_G$, having $|v_i - \langle v \rangle| < 1000 \rm km/s$ and 
$r<17.77$. The properties of these 20 groups are shown in Fig.~\ref{f8} as grey histograms. 
We have checked that the properties of the 2MCGs that have or do not have 
neighbours in SDSS do not present differences.
As previously reported in \cite{diaz12}, the 2MCG 
are typically biased towards lower memberships, fainter surface brightness, 
larger angular diameters and separations between the two brightest galaxies 
compared to mock CGs (see Fig.~\ref{f1}). 

To compare with the semi-analytical predictions, we used the sample of CGs2 defined in the previous section, 
and also restricted the sample to groups with radial velocity 
less than $10000 \, \rm km/s $ to match the observations. 
We applied the same conditions to the mock CGs 
as in the observations regarding the magnitude of the brightest galaxy and restricted the neighbours 
in the mock CGs within the same magnitude range. 
We found 72 CGs2 that satisfy that $r_{brightest} \le 13.27$, $v_r<10000 \, \rm km/s$ and have 
$694$ faint neighbours brighter than $r=17.77$. 

Given the low number of faint galaxies in the observational sample, 
introducing the absolute magnitude limit $M \ge -17$ to define faint galaxies would reduce the number of objects, 
affecting the statistical significance of the results. 
Therefore, we changed the criterion of magnitudes to select faint neighbours, 
allowing the sample to contain all galaxies with apparent magnitudes brighter than $17.77$,
 and excluding only the compact group members ($r_{brightest} \le r \le r_{brightest}+3$).  
However, working with flux limited catalogues implies that for different values of $r_{brightest}$,
a different range of luminosities will be included in the search for faint galaxies.
Then, around apparently bright galaxies (i.e. with bright apparent magnitudes),
the range of faint galaxies included in the profiles will be larger than 
the corresponding range for galaxies that were not as bright in apparent magnitude, 
despite their intrinsic luminosity. Hence, we checked whether this change in the definition
of faint neighbours introduces differences in the results previously found 
for volume limited samples for the semi-analytical galaxies. 
In Fig.~\ref{ff}, we show the number density profiles of 
faint galaxies selected in flux limited samples ($r_{lim}=17.77$) around simulated compact and control groups. 
The samples of compact and control groups are those defined as CGs2 and S2 in Sect.~\ref{profiles}, 
having the same distribution of normalisation sizes.
It can be seen that the ratios between the profiles around CGs control groups 
are consistent with the results obtained for volume-limited samples 
extracted from a deeper catalogue ($r_{lim}=21$, see Fig.~\ref{f7}), 
i.e., the distributions of faint galaxies around compact and normal groups are alike. 
This test allowed us to conclude that provided the samples of faint neighbours around 
compact and control groups are selected under the same restrictions (flux- or volume-limited), 
the comparison between compact and control groups gives similar results.
Thus, in order to compare observations with simulations, we applied this flux-limited criterion 
to select faint galaxies. 
\begin{figure}
\begin{center}
\includegraphics[width=7.0cm]{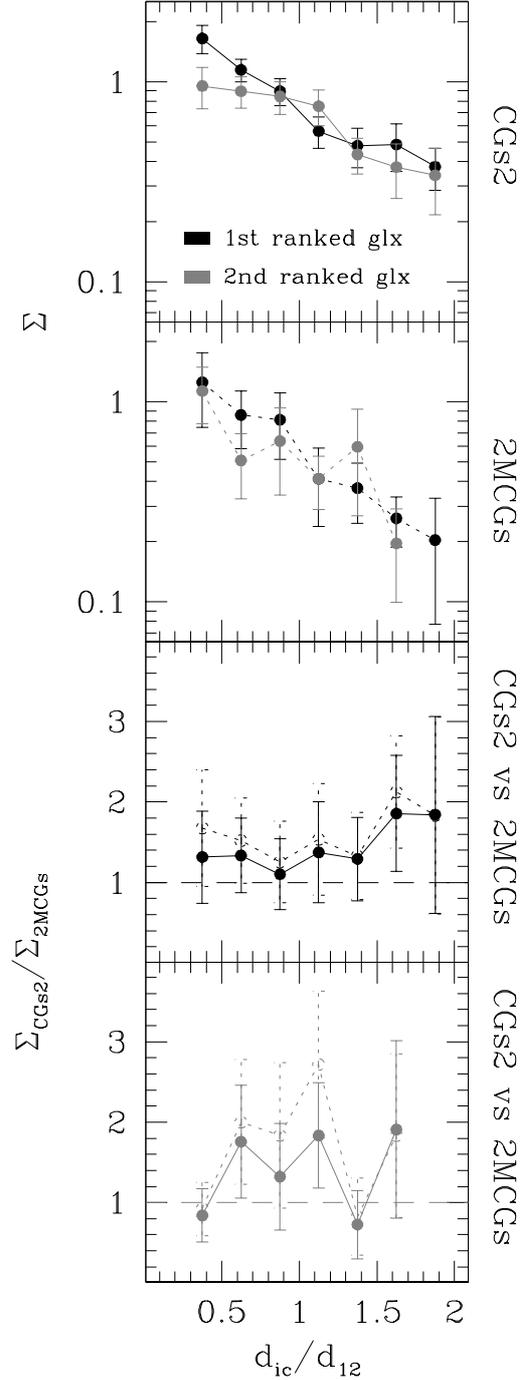}
\caption{Projected number density profiles of faint galaxies around the first ranked
(\emph{black}) and second ranked galaxy of the groups (\emph{grey}).  
The \emph{upper panel} shows the profiles 
for the mock compact group (CGs2) sample 
when incompleteness by fibre collision and low surface brightness are included,
while the \emph{second} panel shows the profiles
for the compact groups in the 2MASS (2MCGs) shown in Fig.~\ref{f8}.
Faint galaxies in the 2MCGs are extracted from the SDSS.
Error bars are the standard deviations computed with 100 bootstraps.
\emph{Bottom panels} show the ratios between the profiles around 
CGs2 and the profiles around the 2MCGs. \emph{Dotted lines} 
shows the corresponding ratios when no incompleteness are included in
the mock sample. Errors are computed by error propagation.
}
\label{f9}
\end{center}
\end{figure}
We included in our sample of mock compact groups two well-known incompletenesses 
present in the SDSS, to allow a fair comparison between observations 
and simulations. 
We first considered the observational limitation of fibre collision in the
spectroscopy of galaxies. Due to restrictions of fibre placement during the SDSS survey, 
two targets separated by less
than $55"$ cannot be observed simultaneously on the same plate, but can be both observed 
on overlapping plates. This fibre collision effect reduces the
number of pairs on small (one-halo) scales and therefore lowers the clustering strength over these
small scales. Instead of correcting the observations by this effect, 
we have modified the mock catalogue by introducing the missing of close pairs of galaxies.
We discarded one galaxy in each pair separated by less than $55"$: 
in the cases where the pair is formed by one galaxy selected as CG member, we discarded
the neighbour; in the cases where the pair is formed by two non-members, we discarded one galaxy 
randomly.
We chose to discard one galaxy in the $100\%$ of those close galaxy pairs, 
which is a higher percentage than in observations, as an extreme case of incompleteness. 
Secondly, there is an incompleteness in the SDSS 
for very low surface brightness objects. The SDSS has 
a Petrosian half-light surface brightness ($\mu_{50}$) limit of 24.5 ${\rm mag/arcsec}^2$ and
becomes incomplete for $\mu_{50}>23 \, {\rm mag/arcsec}^2$.   
Even though this incompleteness is marginal in the range of magnitudes of galaxies involved
in our work, we considered this effect in the semi-analytical faint galaxies.
Since in the semi-analytical model the galaxies are point particles, we assigned 
surface brightness to each galaxy using the empirical prescriptions given by \cite{shen03} 
which relate the absolute magnitude in the $r_{SDSS}$-band, $M_r$, with the half-light radius 
in physical units, $R_{50}$, and the surface brightness with $M_r$ and $R_{50}$
(eqs.~14, 15 and 20 in \citeauthor{shen03}). Using these estimates we discarded 
in the mock faint galaxy sample all galaxies with $\mu_{50}>23 \, {\rm mag/arcsec}^2$.
Therefore, our main mock faint galaxy sample is $100\%$ incomplete due to the missing pair
problem and $100\%$ incomplete for low surface brightness galaxies.
Hence, under these restrictions, we found 69 CGs2 with $555$ faint neighbours.

Figure~\ref{f8} shows a comparison of the properties of the 69 CGs2 (\emph{black lines}) 
and the 20 2MCGs that have faint neighbours around them in the spectroscopic SDSS sample
(\emph{grey histograms}). Both samples span similar ranges in the 
distribution of the group properties.

Regarding the projected number density profiles, we followed the same procedure as explained 
in Sect.~\ref{profiles}. 
The number of galaxies that are effectively taken into account to compute the profiles
are 83(58) around the 1st(2nd) ranked galaxies around 19(18) observable 2MCGs, and
393(337) faint galaxies around the 1st(2nd) ranked galaxies of 64(66) mock CGs.    
Fig.~\ref{f9} shows the profiles around the 1st and 2nd ranked galaxies for mock CGs 
(\emph{solid lines}) and the 2MCGs (\emph{dotted lines}). 
Due to the low number of observational CGs (and faint galaxies around them), we 
show in this figure only those bins of normalised distance in which at least 2 observational
galaxies contribute. This choice restricted the profiles within $\sim 2 \, d_{12}$ from the central galaxies.  
From the ratios among the profiles
shown in the two bottom panels (\emph{solid lines}), it can be seen that, 
in spite of the large error bars resulting due to the small number of galaxies involved,
the distributions of the populations of faint galaxies around mock and observed CGs are 
statistically indistinguishable.   

However, these results are a lower limit to the projected number density profiles given that 
we have overestimated the effect of the fibre collisions and the loss of low surface brightness 
galaxies. We also analysed the scenario where none of these effects are taken into account in the
sample of  faint semi-analytical galaxy neighbours.
The results can be seen in the two bottom panels of Fig.~\ref{f9} where \emph{dotted lines}
represent the ratios between the observed profiles and the simulated profiles without 
introducing any incompleteness. These results are an upper limit to the ratios between the
simulated and observed projected number density profiles. Since the resulting ratios in the
lower and upper limits are indistinguishable within the errors, we conclude that these 
observational incompleteness are not relevant for our comparison.

\section{Summary}
By definition, only galaxies within a 3-magnitude
range from the brightest galaxy of each group are considered as members of CGs.
Isolation and compactness are defined based on the galaxy members, 
therefore fainter galaxies do not affect them. However, they do inhabit the 
same environment as the brighter galaxies in compact groups and 
might feel the effect of an overdense environment in both, 
their abundance and their properties. In this work, 
we focused on the existence and distribution of faint galaxies 
in/around compact groups.
 
To assess the question of whether the CG extreme environment affects the 
abundance of fainter galaxies, 
in this work we explored the projected number density 
profiles of the fainter population of galaxies in these systems. We compared
our results with the profiles of the same population inhabiting normal groups.  

Observationally, the study of the faint population of galaxies in compact 
groups has been limited given the difficulties for detecting such galaxies. 
We faced the problem from a semi-analytical point of view by exploring the
behaviour of the faint population of galaxies in the surroundings of groups
extracted from mock catalogues built from synthetic galaxies extracted 
from a semi-analytical model of galaxy formation run on top of a numerical 
N-body dark matter simulation. 

We chose the publicly available outputs of the \cite{guo11}'s 
semi-analytical model 
combined with the Millennium II simulations \citep{mII} to construct a 
lightcone of galaxies with observable properties of galaxies. 
This model has been tuned to match the observable properties of galaxies at redshift zero, 
with particular emphasis on the faint galaxy population which makes it the most suitable 
to perform the analyses developed in this present work.
 
Compact groups were identified following the standard criteria established
by \cite{hickson82} modified to reproduce the largest and more complete sample of 
compact groups that have been identified automatically from the 2MASS catalogue \citep{diaz12}.
Normal groups were identified using the standard Friends-of-Friends algorithm applied to a flux 
limited catalogue \citep{huchra82,fof14}.

We computed the projected number density profiles of faint galaxies in/around the main galaxies 
of groups and, in order to stack groups, we used as normalisation parameter the separation 
between the first- and second-ranked galaxies of the groups.  

Firstly we observe that, the shape of the projected number density profiles in the inner 
regions indicates that the faint galaxy population is more concentrated around the first ranked
galaxy than around the second ranked galaxy, for CG and control samples. 
We found that compact groups are underdense in faint galaxies when compared to normal groups, 
however, the shape of the distribution of the faint populations around both types of systems is alike.
 
Given that one of the main differences between compact groups and normal groups 
is the size of the normalisation parameter, we computed the profiles for subsamples 
of compact and normal groups with the same distribution of the normalisation sizes. 
This time, the projected number density profiles around the first ranked galaxies 
of compact an normal groups look alike, not only in shape but also in height, 
indicating that there is no particular influence of the extreme compact 
group environment on the number of faint galaxies in such groups.  

We also compared the distribution of the faint population around semi-analytical and observational 
compact groups. We used the observational compact groups identified by \cite{diaz12} that lie on 
the SDSS area, from where we extracted the fainter galaxies with spectroscopic information. 
Although the number of observational compact groups and faint neighbours is small, 
this exercise allowed us to compare the predictions of the semi-analytical model to observations.
We observed a similar behaviour of the population of faint galaxies in observations and simulations
with the semi-analytical model of \cite{guo11}. 
Different authors have performed similar comparisons but using different semi-analytical models 
and different types of groups. \cite{weinmann06,liu10} found that previous versions of the semi-analytical 
models \citep{croton06,bower06,kang05} overpredicted the satellite content of groups and clusters. However, 
\cite{weinmann11} found that the number density profile
of faint galaxies (in a fixed absolute magnitude range) in massive clusters is well reproduced by 
the state-of-the-art semi-analytical models \citep{guo11}. 
In this work, we were able to extend this result to compact groups.
Nevertheless, we note the need for more observational data to perform a more reliable 
comparison.   

\begin{acknowledgements}
We acknowledge the anonymous referee for insightful comments and suggestions 
that increased the general quality of the paper.

The Millennium Simulation databases used in this paper and the web application providing online 
access to them were constructed as part of the activities of the German Astrophysical 
Virtual Observatory (GAVO).
We thank Qi Guo for allowing public access for the outputs of her very impressive
semi-analytical model of galaxy formation.

Funding for SDSS-III has been provided by the Alfred P. Sloan Foundation, the Participating Institutions, 
the National Science Foundation, and the U.S. Department of Energy Office of Science. 
The SDSS-III web site is http://www.sdss3.org/.
SDSS-III is managed by the Astrophysical Research Consortium for the Participating Institutions of the SDSS-III Collaboration including the University of Arizona, the Brazilian Participation Group, Brookhaven National Laboratory, Carnegie Mellon University, University of Florida, the French Participation Group, the German Participation Group, Harvard University, the Instituto de Astrofisica de Canarias, the Michigan State/Notre Dame/JINA Participation Group, Johns Hopkins University, Lawrence Berkeley National Laboratory, Max Planck Institute for Astrophysics, Max Planck Institute for Extraterrestrial Physics, New Mexico State University, New York University, Ohio State University, Pennsylvania State University, University of Portsmouth, Princeton University, the Spanish Participation Group, University of Tokyo, University of Utah, Vanderbilt University, University of Virginia, University of Washington, and Yale University.

This work has been partially supported by Consejo Nacional de Investigaciones Cient\'\i ficas y 
T\'ecnicas de la Rep\'ublica Argentina (CONICET), Secretar\'\i a de 
Ciencia y Tecnolog\'\i a de la Universidad de C\'ordoba (SeCyT) and Funda\c c\~ao de Amparo \`a 
Pesquisa do Estado do S\~ao Paulo (FAPESP), through grants 2011/50471-4 and 2011/50002-4. CMdO 
acknowledges support of FAPESP (grant \#2006/56213-9) and Conselho Nacional de Pesquisas (CNPq). 
HG thanks FAPESP for the IC grant \#2012/04106-5.
\end{acknowledgements}

\bibliography{refs}

\appendix
\section{Projected number density profiles for different classes of Compact Groups: 
ranking vs luminosity}
\label{appen1}

\begin{figure*}
\begin{center}
\includegraphics[width=13cm]{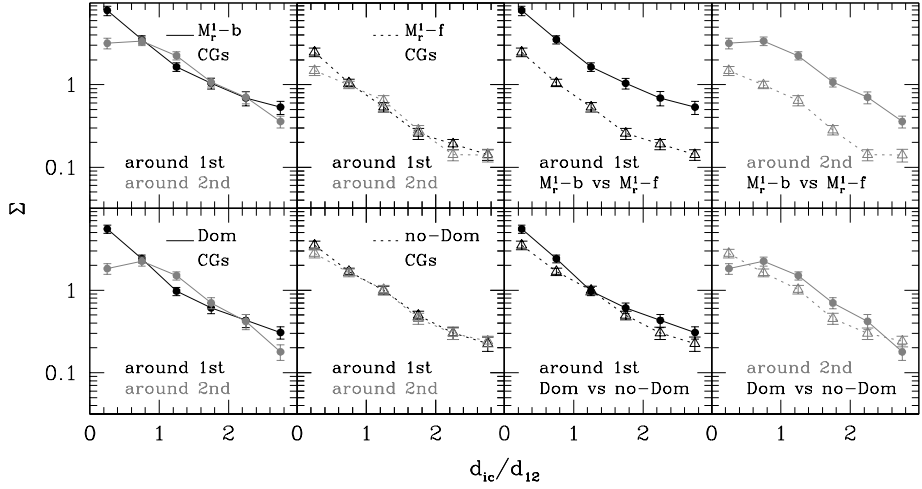}
\caption{Projected number density profiles of faint galaxies around the first ranked
(\emph{black}) and second ranked galaxy (\emph{grey}) for different subsamples of Compact Groups.
Faint galaxies are defined within a volume limited sample with $-17 \le \rm M_r - 5\log{(h)} \le -13.8$ 
and $z_{cm}<0.03$.
\emph{Top panels}: compact groups with bright/faint first-ranked galaxies. \emph{Bottom panels}: compact groups 
Dominated/non-Dominated by a single bright galaxy.   
Error bars are the standard deviations computed with 100 bootstraps.
}
\label{fa1}
\end{center}
\end{figure*}

We investigated what is the most important parameter in the determination of the centre around which 
the clustering of faint galaxies occurs in compact groups, i.e. if faint galaxies 
are preferentially distributed around the most luminous galaxy 
(in which case the most important is the ranking of the galaxy) or around any 
luminous galaxy (most important parameter is the luminosity). 
Therefore, we studied the density profiles of faint galaxies
around different subsamples of CGs classified according to different
group properties. The criteria used to split the 
subsamples are:

\begin{itemize}
\item Faint vs Bright first-ranked galaxy: we split the sample
of CGs according to the absolute magnitude of the brightest
galaxy of the system. Those having a brightest galaxy
brighter than the 30-th percentile of the distribution of absolute magnitudes of 
the 1st-ranked galaxies of all the CGs are classified as $M^1_r-b$, 
while CGs whose first-ranked galaxy is fainter 
than the 70-th percentile of the distribution of absolute magnitudes 
of the first-ranked galaxies in the complete sample of CGs are $M^1_r-f$.
\item Dominated vs non-Dominated groups: we split the sample
of CGs according to the difference in absolute
magnitude between the first- and second-ranked galaxies.
CGs dominated by a very bright galaxy (Dom) will exhibit a larger
difference between their two brightest members. We used the 30-th and 70-th percentiles of the 
distribution to split the sample into "Dom" and "non-Dom" CGs, respectively.
\end{itemize}

Faint galaxies are selected from a volume limited sample having 
$-17 \le \rm M_r - 5\, \log{(h)} \le -13.8$ and $z_{cm} \le 0.03$ (see Fig.~\ref{f4}). 
Figure~\ref{fa1} shows the projected number density profiles of faint
galaxies around different classes of CGs. It can be seen that:
\begin{itemize}
\item Groups having the brighter first-ranked galaxies 
are denser around the 1st- and the 2nd-ranked galaxies 
(\emph{top panels}).
\item The projected number density profiles around the 1st- and 2nd-ranked galaxies 
of CGs non-dominated by a single galaxy are remarkably similar.   
\end{itemize}

These two results lead us to conclude that faint galaxies are more frequent around bright galaxies, 
regardless the ranking in the CG.

\end{document}